\documentclass[twocolumn,trackchanges]{aastex701}

\newcommand{\hi}{\ion{H}{1}}
\newcommand{\kms}{\ensuremath{{\rm km~s^{-1}}}}
\newcommand{\persc}{\ensuremath{{\rm cm^{-2}}}}
\newcommand{\percc}{\ensuremath{{\rm cm^{-3}}}}

\begin{document}

\shortauthors{LGLBS Collaboration}

\title{Revisiting ram pressure stripping in Wolf-Lundmark-Melotte: No evidence for stripped \hi\ with LGLBS}

\newcommand{\Ox}{Sub-department of Astrophysics, Department of Physics, University of Oxford, Keble Road, Oxford OX1 3RH, UK}

\newcommand{\UGent}{Sterrenkundig Observatorium, Universiteit Gent, Krijgslaan 281 S9, B-9000 Gent, Belgium}

\newcommand{\STScI}{Space Telescope Science Institute, 3700 San Martin Drive, Baltimore, MD 21218, USA}

\newcommand{\IWR}{Universit\"{a}t Heidelberg, Interdisziplin\"{a}res Zentrum f\"{u}r Wissenschaftliches Rechnen, Im Neuenheimer Feld 205, 69120 Heidelberg, Germany}
\newcommand{\Radcliffe}{Elizabeth S. and Richard M. Cashin Fellow at the Radcliffe Institute for Advanced Studies at Harvard University, 10 Garden Street, Cambridge, MA 02138, U.S.A.}

\newcommand{\MPIA}{Max-Planck-Institut f\"{u}r Astronomie, K\"{o}nigstuhl 17, D-69117, Heidelberg, Germany}

\newcommand{\AURA}{AURA for the European Space Agency (ESA), Space Telescope Science Institute, 3700 San Martin Drive, Baltimore, MD 21218, USA}

\newcommand{\UCSD}{Department of Astronomy \& Astrophysics, University of California, San Diego, 9500 Gilman Dr., La Jolla, CA 92093, USA}

\newcommand{\Whitman}{Whitman College, 345 Boyer Avenue, Walla Walla, WA 99362, USA}

\newcommand{\JHU}{Department of Physics and Astronomy, The Johns Hopkins University, Baltimore, MD 21218, USA}

\newcommand{\OSU}{Department of Astronomy, The Ohio State University, 140 West 18th Avenue, Columbus, OH 43210, USA}

\newcommand{\OSUPhys}{Department of Physics, The Ohio State University, Columbus, Ohio 43210, USA}

\newcommand{\CCAPP}{Center for Cosmology and Astroparticle Physics (CCAPP), 191 West Woodruff Avenue, Columbus, OH 43210, USA}

\newcommand{\ARI}{Astronomisches Rechen-Institut, Zentrum f\"{u}r Astronomie der Universit\"{a}t Heidelberg, M\"{o}nchhofstr. 12-14, D-69120 Heidelberg, Germany}

\newcommand{\ANU}{Research School of Astronomy and Astrophysics, Australian National University, Canberra, ACT 2611, Australia}

\newcommand{\UConn}{Department of Physics, University of Connecticut, 196A Auditorium Road, Storrs, CT 06269, USA}

\newcommand{\UHawaii}{Institute for Astronomy, University of Hawaii, 2680 Woodlawn Drive, Honolulu, HI 96822, USA}

\newcommand{\UniCA}{Universit\'{e} C\^{o}te d'Azur, Observatoire de la C\^{o}te d'Azur, CNRS, Laboratoire Lagrange, 06000, Nice, France}

\newcommand{\UAlberta}{Dept. of Physics, University of Alberta, 4-183 CCIS, Edmonton, Alberta, T6G 2E1, Canada}

\newcommand{\Arcetri}{INAF — Osservatorio Astrofisico di Arcetri, Largo E. Fermi 5, I-50125, Florence, Italy}

\newcommand{\UWyoming}{Department of Physics and Astronomy, University of Wyoming, Laramie, WY 82071, USA}

\newcommand{\LJMU}{Astrophysics Research Institute, Liverpool John Moores University, 146 Brownlow Hill, Liverpool L3 5RF, UK}

\newcommand{\ITA}{Universit\"{a}t Heidelberg, Zentrum f\"{u}r Astronomie, Institut f\"{u}r Theoretische Astrophysik, Albert-Ueberle-Str 2, D-69120 Heidelberg, Germany}

\newcommand{\CfA}{Center for Astrophysics $\mid$ Harvard \& Smithsonian, 60 Garden St., 02138 Cambridge, MA, USA}

\newcommand{\MPE}{Max-Planck-Institut f\"{u}r Extraterrestrische Physik (MPE), Giessenbachstr. 1, D-85748 Garching, Germany}

\newcommand{\UMD}{Department of Astronomy and Joint Space-Science Institute, University of Maryland, College Park, MD 20742, USA}

\newcommand{\UVA}{Department of Astronomy, University of Virginia, Charlottesville, VA, USA}

\newcommand{\NRAO}{National Radio Astronomy Observatory, Charlottesville, VA, USA}

\newcommand{\ASIAA}{Institute of Astronomy and Astrophysics, Academia Sinica, No. 1, Sec. 4, Roosevelt Road, Taipei 106216, Taiwan}

\newcommand{\kipac}{Kavli Institute for Particle Astrophysics \& Cosmology (KIPAC), Stanford University, CA 94305, USA}

\newcommand{\aifa}{Argelander-Institut f\"ur Astronomie, University of Bonn, Auf dem H\"ugel 71, 53121 Bonn, Germany}

\newcommand{\TKU}{Department of Physics, Tamkang University, No.151, Yingzhuan Road, Tamsui District, New Taipei City 251301, Taiwan}

\newcommand{\CarnegieObs}{The Observatories of the Carnegie Institution for Science. 813 Santa Barbara Street, Pasadena, CA 91101, USA}

\newcommand{\Princeton}{Department of Astrophysical Sciences, Princeton University, Princeton, NJ 08544, USA}

\newcommand{\IAS}{Institute for Advanced Study, 1 Einstein Drive, Princeton, NJ 08540, USA}

\newcommand{\COOL}{Cosmic Origins Of Life (COOL) Research DAO, coolresearch.io}

\newcommand{\ESO}{European Southern Observatory (ESO), Karl-Schwarzschild-Stra{\ss}e 2, 85748 Garching, Germany}

\newcommand{\ULyon}{Univ Lyon, Univ Lyon 1, ENS de Lyon, CNRS, Centre de Recherche Astrophysique de Lyon UMR5574, F-69230 Saint-Genis-Laval, France}

\newcommand{\UoM}{Jodrell Bank Centre for Astrophysics, Department of Physics and Astronomy, University of Manchester, Oxford Road, Manchester M13 9PL, UK}

\author[0000-0003-3351-6831]{Daniel R. Rybarczyk}
\affiliation{University of Wisconsin–Madison, Department of Astronomy, 475 N Charter St, Madison, WI 53703, USA}
\email{rybarczyk@astro.wisc.edu}

\author[0000-0001-9605-780X]{Eric W. Koch}
\affiliation{National Radio Astronomy Observatory, 800 Bradbury SE, Suite 235, Albuquerque, NM 87106}
\email{koch.eric.w@gmail.com}

\author[0009-0002-3887-5091]{Fabian Caballero Vargas}
\affiliation{New Mexico Institute of Mining and Technology, Physics Department, 801 Leroy Place, Socorro, NM 87801, USA}
\email{fabian.caballerovargas@student.nmt.edu}

\author[0000-0002-3418-7817]{Sne{\v z}ana Stanimirovi{\'c}}
\affiliation{University of Wisconsin–Madison, Department of Astronomy, 475 N Charter St, Madison, WI 53703, USA}
\email{sstanimi@astro.wisc.edu}

\author[0000-0001-9504-7386]{Nickolas~M.~Pingel}
\affiliation{University of Wisconsin–Madison, Department of Astronomy, 475 N Charter St, Madison, WI 53703, USA}
\affiliation{Indiana University, Department of Astronomy, 727 East Third Street, Bloomington, Indiana 47405, USA}
\email{nmpingel@wisc.edu}

\author[0000-0002-1264-2006]{Julianne J.~Dalcanton}
\affiliation{Center for Computational Astrophysics, Flatiron Institute, 162 Fifth Avenue, New York, NY 10010, USA}
\affiliation{Department of Astronomy, Box 351580, University of Washington, Seattle, WA 98195, USA}
\email{jdalcanton@flatironinstitute.org}

\author[0000-0002-2545-1700]{Adam~K.~Leroy}
\affiliation{\OSU}
\affiliation{\CCAPP}
\email{leroy.42@osu.edu}

\author[0000-0002-5204-2259]{Erik~W.~Rosolowsky}
\affiliation{\UAlberta}
\email{rosolowsky@ualberta.ca}

\author[0000-0003-4961-6511]{Michael P. Busch}
\altaffiliation{Jansky Fellow of the National Radio Astronomy Observatory}
\affiliation{National Radio Astronomy Observatory, 520 Edgemont Road, Charlottesville, VA 22903, USA}
\email{mpbusch@nrao.edu}

\author[0000-0003-2896-3725]{Chang-Goo Kim}
\affiliation{Department of Astrophysical Sciences, Princeton University, 4 Ivy Lane, Princeton, NJ 08544, USA}
\email{changgoo@princeton.edu}

\author[0000-0003-2599-7524]{Adam Smercina}\thanks{Hubble Fellow}
\affiliation{Space Telescope Science Institute, 3700 San Martin Dr., Baltimore, MD 21218, USA}
\email{asmerci@uw.edu}

\author[0000-0003-1356-1096]{Elizabeth Tarantino} \affiliation{Space Telescope Science Institute, 3700 San Martin Dr., Baltimore, MD 21218, USA}
\email{etarantino@stsci.edu}

\author[0000-0002-5877-379X]{Vicente Villanueva}
\affiliation{Departamento de Astronom\'ia, Universidad de Concepci\'on, Barrio Universitario, Concepci\'on, Chile}
\email{vvillanueva@astro-udec.cl}

\author[0000-0002-5480-5686]{Alberto D. Bolatto}
\affiliation{Department of Astronomy and Joint Space-Science Institute, University of Maryland, College Park, MD 20742, USA}
\email{bolatto@umd.edu}

\author[0000-0002-0012-2142]{Thomas~G.~Williams}
\affiliation{UK ALMA Regional Centre Node, Jodrell Bank Centre for Astrophysics, Department of Physics and Astronomy, The University of Manchester, Oxford Road, Manchester M13 9PL, UK}
\email{thomas.g.williams@manchester.ac.uk}

\begin{abstract}

We analyze \ion{H}{1} 21-cm observations of the Local Group dwarf galaxy Wolf-Lundmark-Melotte (WLM) from the Local Group L-Band Survey to search for evidence of ram pressure stripping. While previous MeerKAT-16 observations of WLM showed evidence for off-galaxy atomic gas emission with a geometry suggestive of ram pressure stripping, our observations find no evidence for this stripped gas. We demonstrate that our observations would be sensitive to the claimed detections and suggest that an uncorrected observational flaw with the MeerKAT data led to the apparent off-galaxy emission. The lack of off-galaxy emission obviates the need for uncharacteristically high values of the density of the intergalactic medium in this region.
\end{abstract}

\section{Introduction}  \label{sec:intro}
Wolf-Lundmark-Melotte (WLM, DDO 221, UGCA 444) is a dwarf irregular galaxy located at a distance of $984\pm16$~kpc \citep{lee2021dist}.
It is highly isolated, with the nearest known neighbor located 200~kpc away \citep{Kepley2007}.  WLM therefore offers an excellent laboratory for studying the structure and kinematics of non-interacting dwarf galaxies and the density of an intragroup medium. 

Recently, \citet{Ianjamasimanana2020} characterized the distribution and kinematics of the atomic hydrogen (\hi) in WLM using 21-cm emission observations from MeerKAT-16 (the early 16-dish configuration of MeerKAT) with an angular resolution of $35\arcsec\times12\arcsec$ (although much of their analysis used a $60\arcsec \times 60\arcsec$ version of the data cube) and a velocity resolution of $5.5~\kms{}$.
Using the observations obtained by \citet{Ianjamasimanana2020}, \citet{Yang2022} presented evidence for diffuse, ram-pressure-stripped \hi\ trailing WLM. Specifically, they identified four \hi\ clouds, at high signal-to-noise (peak S/N $\gtrsim20$), positioned $\sim10$ to 20 arcminutes northwest of the main body of the galaxy. 
\citet{Kolhe2026} re-analyzed the \citet{Ianjamasimanana2020} data and found a smoother distribution of trailing gas rather than four discrete clouds.
\citet{Yang2022} argued that the spatial distribution of these clouds --- opposite to WLM's direction of motion --- was evidence for ram pressure stripping in this isolated dwarf galaxy, and that about 10\% of the total \hi\ mass was stripped by ram pressure.
However, new deeper \hi\ observations using the full MeerKAT array from \cite{Kolhe2026} did not detect emission from such trailing gas, which they argue results from less short-baseline coverage compared to \citet{Ianjamasimanana2020}. 

\citet{Koch2025} presented WLM observations taken with the Karl G. Jansky Very Large Array (VLA) as part of the Local Group L-Band Survey (LGLBS), which measured 21-cm emission with high sensitivity and resolution (both spatial and spectral) toward six Local Group galaxies.
These data sample similar scales to both MeerKAT-16 and full array observations and are $4\times$ more sensitive than the MeerKAT-16 observations on $\sim1\arcmin$ scales.  These observations should therefore detect the same features found in the MeerKat data, if present.

The detection of a ram-pressure-stripped tail of gas from an isolated galaxy in the Local Group has important implications on the conditions required to quench galaxies.  The interplay of the interstellar medium (ISM), circumgalactic medium (CGM), and intergalactic medium (IGM) as a dwarf galaxy moves through a diffuse surrounding medium has been studied in both observation and simulation work, and in general there is agreement on the overall conditions required \citep[e.g.,][]{ZhuPutman2023,HumaranMinchin2025,Luber2025}.  In contrast, the \citet{Yang2022} claim of stripped clouds from WLM imply an IGM density significantly higher than that derived from previous observations of the dwarf galaxies Holmberg II \citep{BureauCarignan2002} and the Local Group Pegasus dwarf irregular \citep{McConnachie2007}.

Using the LGLBS data, we search for, but do not find, evidence of ram-pressure-stripped \hi\ in and around WLM.

\section{Revisiting ram pressure stripping with LGLBS} \label{sec:lglbs}

\begin{figure*}
    \centering
    \includegraphics[width=0.57\linewidth]{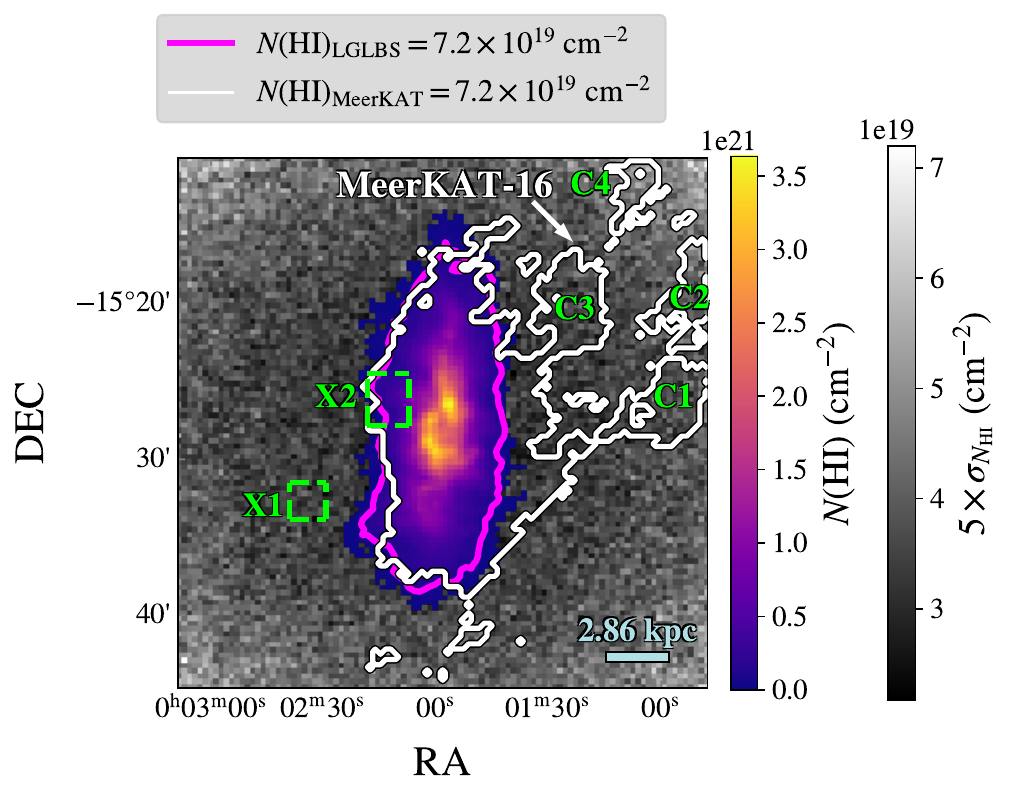}
    \includegraphics[width=0.41\linewidth]{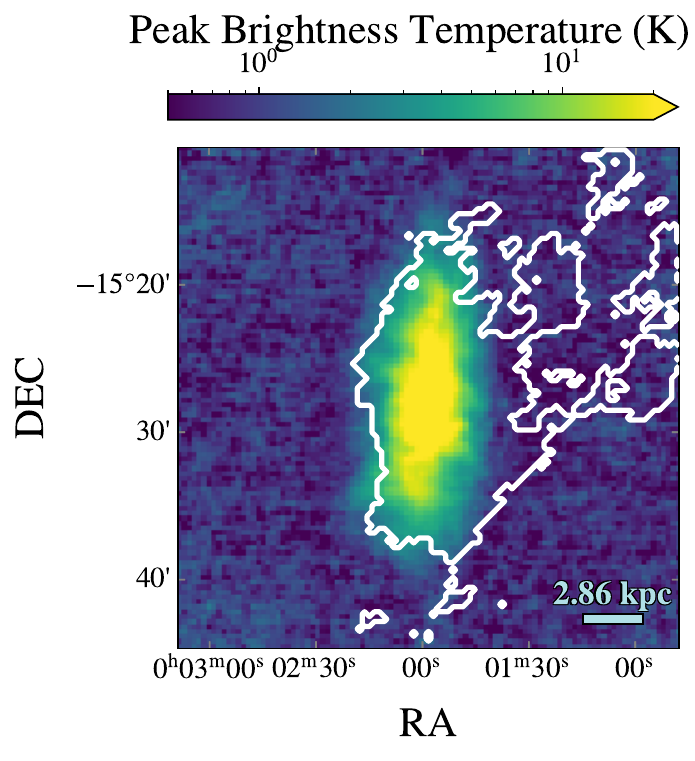}
    \includegraphics[width=\linewidth]{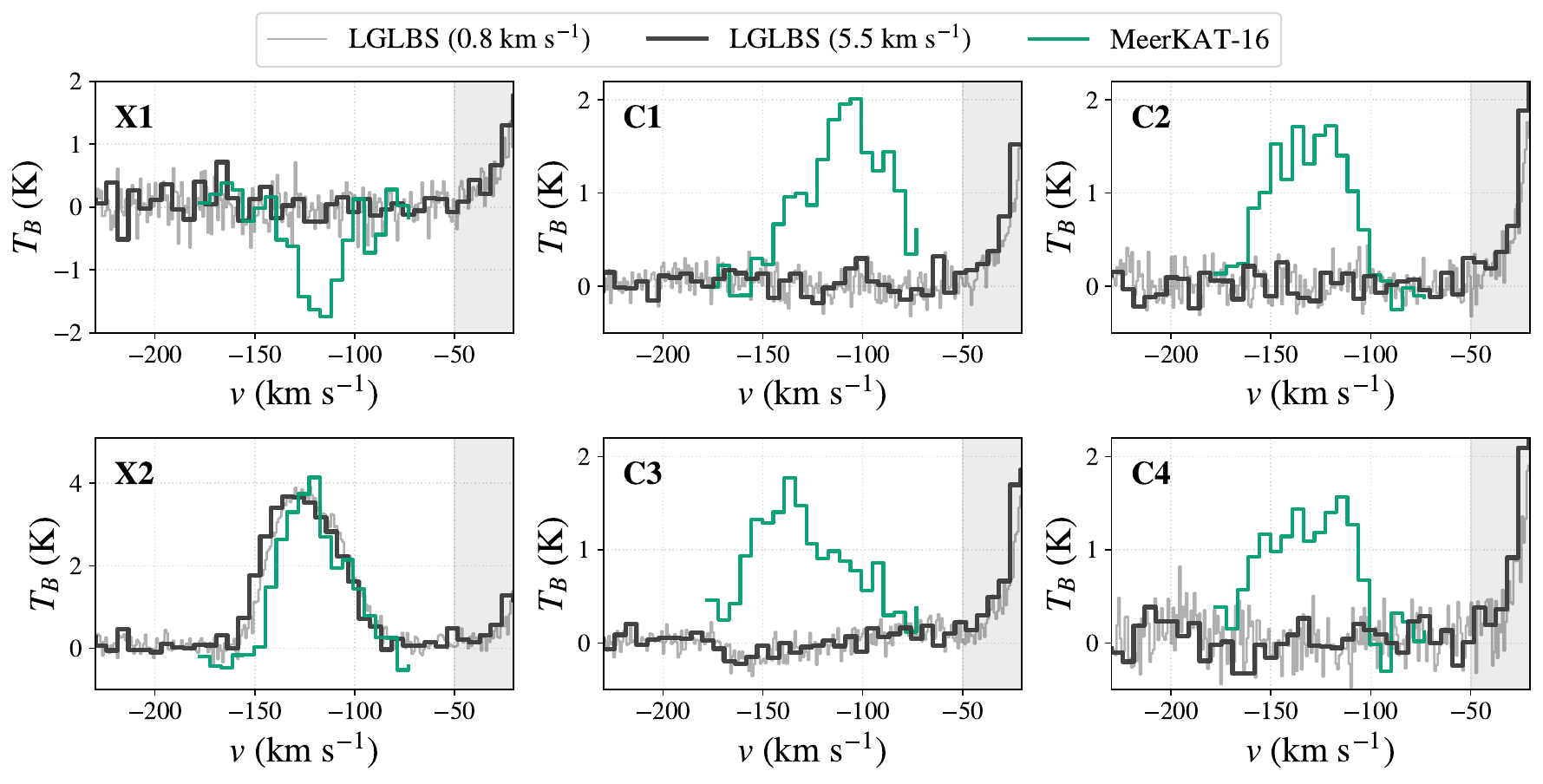}
    \caption{\textit{Top left}: LGLBS WLM \hi\ column density map at $60\arcsec$ resolution. For pixels without significant \hi\ emission, the background grayscale indicates the $5\sigma$ column density sensitivity. Contours are shown at 
    $N(\text{\ion{H}{1}})=7.2\times10^{19}~\persc$ --- corresponding to the $3\sigma$ sensitivity for the MeerKAT-16 observations \citep{Ianjamasimanana2020} --- for both the LGLBS data (magenta) and the MeerKAT-16 data \citep[white; ][]{Ianjamasimanana2020,Yang2022}. The clouds identified by \citet{Yang2022} in the MeerKAT-16 data are labelled C1, C2, C3, and C4. The two regions where we extract spectra for controls are labelled X1 and X2. 
    \textit{Top right:} Peak brightness temperature \citep[over $30~\kms$ channels, the approximate width of the emission reported by][]{Yang2022} in the LGLBS data. 
    The MeerKAT $N(\text{\ion{H}{1}})=7.2\times10^{19}~\persc$ contour is shown in white.
    \textit{Bottom:} Average \hi\ emission spectra from the $60\arcsec$ LGLBS data (at $0.8~\kms$ velocity resolution in semi-transparent black and at $5.5~\kms$ velocity resolution in solid black) and the $60\arcsec$ MeerKAT-16 data (green; $5.5~\kms$ velocity resolution), averaged over the regions identified in the top panel --- C1, C2, C3, C4, X1, and X2. The gray shaded regions of the spectra indicate the velocity range of Milky Way \hi\ emission.
    Despite our improved sensitivity, we do not find evidence of the ram-pressure-stripped \hi\ reported by \citet{Yang2022}. Meanwhile, toward the X1 region, we find evidence of a background that is not flat across the \citet{Ianjamasimanana2020} \hi\ cube.}
    \label{fig:LGLBS_MeerKAT_comparison}
\end{figure*}

To make a direct comparison to the previously-reported detection of ram-pressure-stripped atomic gas, we here use a version of the LGLBS \hi\ data cube made with a $60\arcsec$ Gaussian beam --- the same resolution as the MeerKAT-16 \hi\ cube presented by \citet{Ianjamasimanana2020} and used in the analysis of \citet{Yang2022}.
The LGLBS \hi\ cube combines VLA C and D configurations with short-spacing correction included by feathering with the GBT.
Detailed information on data handling and imaging are given in \citet{Koch2025}. 
At the matched $60\arcsec$ resolution, the LGLBS data have a $5\sigma$ \hi\ column density sensitivity of $2.8 \times 10^{19}~\persc$ per $10~\kms$, which is $\sim4\times$ better sensitivity than the MeerKAT-16 data \citep[$1.2 \times 10^{20}~\persc$ per $10~\kms$;][]{Ianjamasimanana2020,Yang2022}.

We do not detect \hi\ emission from ram-pressure-stripped gas in the LGLBS data despite achieving better sensitivity than the \citet{Ianjamasimanana2020} MeerKAT-16 observations. 
In Figure \ref{fig:LGLBS_MeerKAT_comparison}, we show the \hi\ column density map 
constructed from the LGLBS data and the $5\sigma$ \hi\ column density sensitivity for pixels without significant emission. Overlaid in white are contours of the \hi\ column density reported by \citet{Yang2022}, tracing the boundary of \hi\ emission detected at $>3\sigma$ in the \citet{Ianjamasimanana2020} observations. The four clouds that \citet{Yang2022} identified as ram-pressure-stripped \hi\ are labelled C1, C2, C3, and C4. They reported typical \hi\ column densities $\sim(1$--$3)\times10^{21}~\persc$ for these clouds. Given that they found S/N $>20$ detections, we should easily see the features in our data cube. 
Instead, with the LGLBS observations, we place upper limits on the column density of any potentially stripped \hi\ roughly an order of magnitude lower than these reported values, $N(\text{\ion{H}{1}})<2.8\times10^{19}~\persc$.
We also show the peak brightness temperature measured by LGLBS in Figure \ref{fig:LGLBS_MeerKAT_comparison}, which is $\lesssim1~$K over virtually the entire region where \citet{Yang2022} identified stripped gas.
Whereas \citet{Yang2022} reported that roughly $10\%$ of the \hi\ mass of WLM was in these stripped gas clouds, we find $<0.1\%$ of WLM's \hi\ mass exists in these directions.

We further compare the \hi\ emission spectra in several key directions in Figure \ref{fig:LGLBS_MeerKAT_comparison}. For each of the four regions where \citet{Yang2022} reported the detection of a ram-pressure-stripped cloud (C1, C2, C3, and C4), we show the average \hi\ spectra from the $60\arcsec$ LGLBS cube (here taken at $0.8~\kms{}$ velocity resolution) in black and the average spectra from the $60\arcsec$ MeerKAT cube (private communication) in green. As shown in \citet{Yang2022}, there appears to be emission in the MeerKAT data at velocities $-160~\kms \lesssim v \lesssim -100~\kms$ at S/N~$\gtrsim 20$. We find no significant emission from WLM in the LGLBS data in these directions at any velocity.

For reference, we also show the spectra over two control regions --- ``X1,'' a region 
$\sim3\arcmin$ from the westernmost emission detected in either the LGLBS or the MeerKAT observations; and ``X2,'' a region where both surveys detected significant emission, slightly offset from the major axis of WLM. Both regions --- comparable in size to the clouds identified by \citet{Yang2022} --- are outlined in Figure \ref{fig:LGLBS_MeerKAT_comparison}. 
The spectra from LGLBS and the MeerKAT-16 observations in 
the X2 control region is in good agreement, demonstrating that the difference in detections for regions C1--C4 are not consistent with other regions throughout the maps. 
Meanwhile, the MeerKAT spectrum toward the X1 control region shows a significant negative trough, similar in amplitude to the detections reported by \citet{Yang2022} toward C1--C4. Such a dip in the emission spectrum could perhaps point to the presence of unflagged radio frequency interference (RFI) and the resultant striping in the emission maps (see discussion in Section \ref{sec:discussion}).

\begin{figure}[!t]
    \centering
    \includegraphics[width=0.48\textwidth]{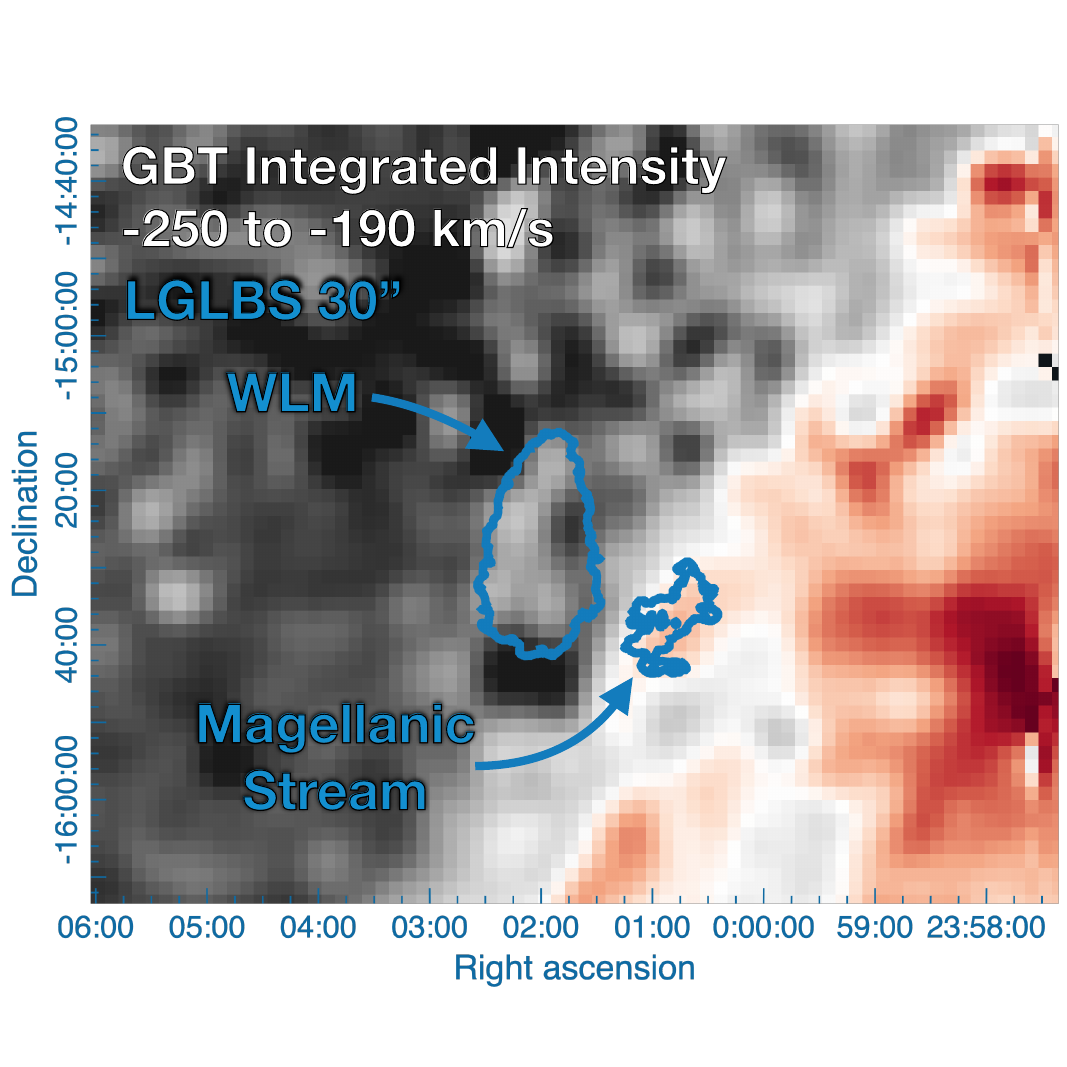}
    \includegraphics[width=0.48\textwidth]{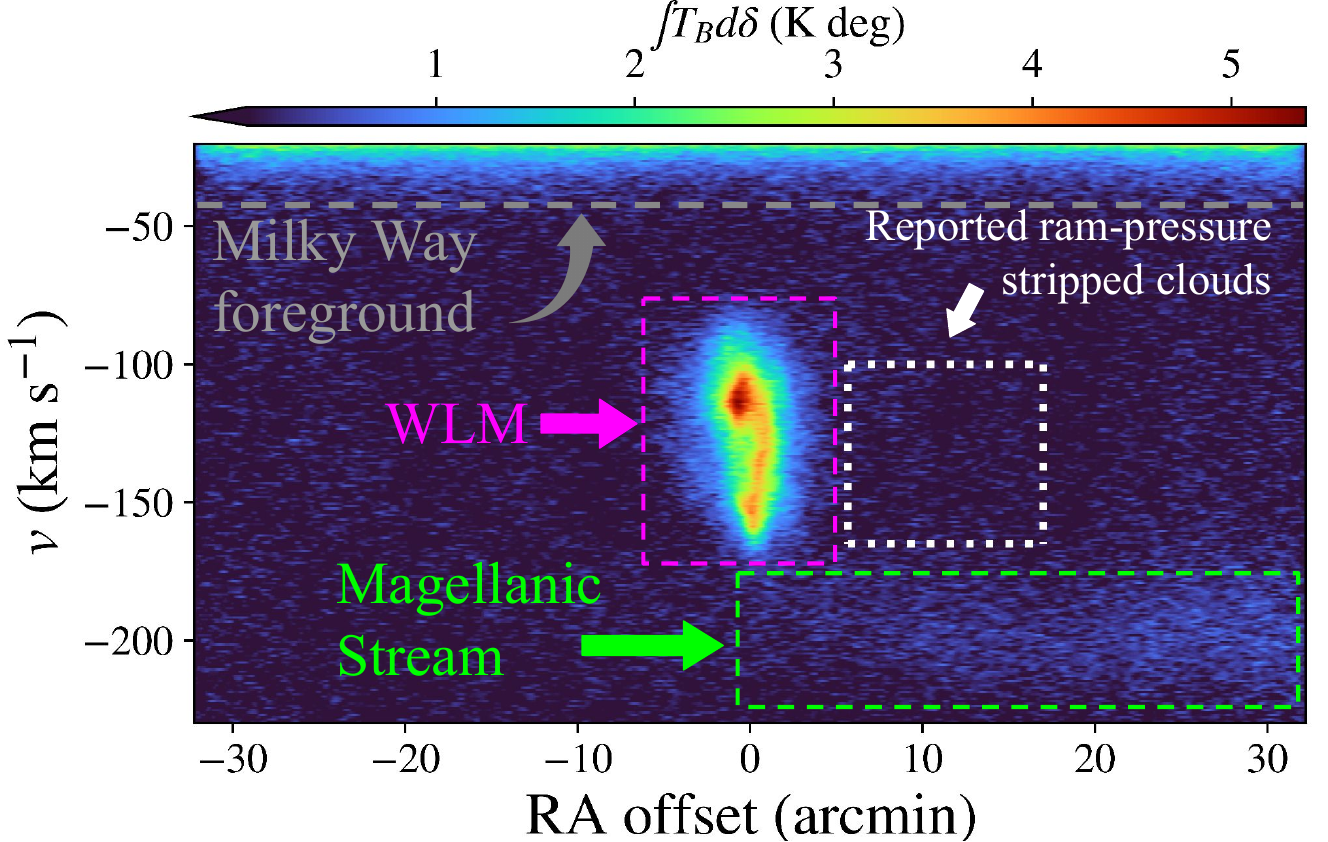}
    \caption{\textit{Top:} GBT integrated intensity from $-190$ to $-250$~\kms\ highlighting \hi\ emission from the Magellanic Stream.
    The blue contour show where \hi\ is detected ($>1$~K~\kms) from the LGLBS \hi\ map when smoothed to $30\arcsec$.
    \textit{Bottom:} Position-velocity diagram of the \hi\ emission measured by LGLBS. Data are integrated over the full declination range observed by LGLBS; the $x$-axis position marks the offset in right ascension from the center of the image (approximately aligned with the major axis of WLM). Contributions from the Milky Way foreground, WLM, and the Magellanic Stream are outlined in dark-gray, magenta, and green dashed boxes, respectively. The RA/velocity range of the ram-pressure-stripped clouds reported by \citet{Yang2022} is outlined in a white dashed box. 
    While we detect the diffuse emission from the Magellanic Stream, we do not detect \hi\ from clouds at the positions and velocities reported by \citet{Yang2022}.
    }
    \label{fig:wlm_magstream_gbt}
\end{figure}

\citet{Kolhe2026} recently presented new MeerKAT-64 observations of WLM obtained with 61 MeerKAT dishes\footnote{We refer to the full MeerKAT array as MeerKAT-64, noting that only 61 dishes were used for the \citet{Kolhe2026} observations. Nonetheless, their observations are representative of the full MeerKAT array relative to early MeerKAT-16 observations.}. Like LGLBS, they did not detect any of the ram-pressure-stripped \hi\ reported by \citet{Yang2022}.
They argued that this non-detection was the result of the 61-dish $uv$ configuration, which they claimed limited the sensitivity to extended emission, consistent with their non-detection of foreground emission from the Magellanic Stream \citep{Putman2003}; the data imaged by \citet{Ianjamasimanana2020} did not have sufficient velocity coverage to search for Magellanic Stream emission.
In the LGLBS data, we recover the \hi\ foreground emission from the Magellanic Stream to the southeast of WLM \citep[Figure \ref{fig:wlm_magstream_gbt}; see][]{Koch2025}, but do not detect the clouds identified by \citet{Yang2022}.

\begin{figure}
    \centering
    \includegraphics[width=1.05\linewidth]{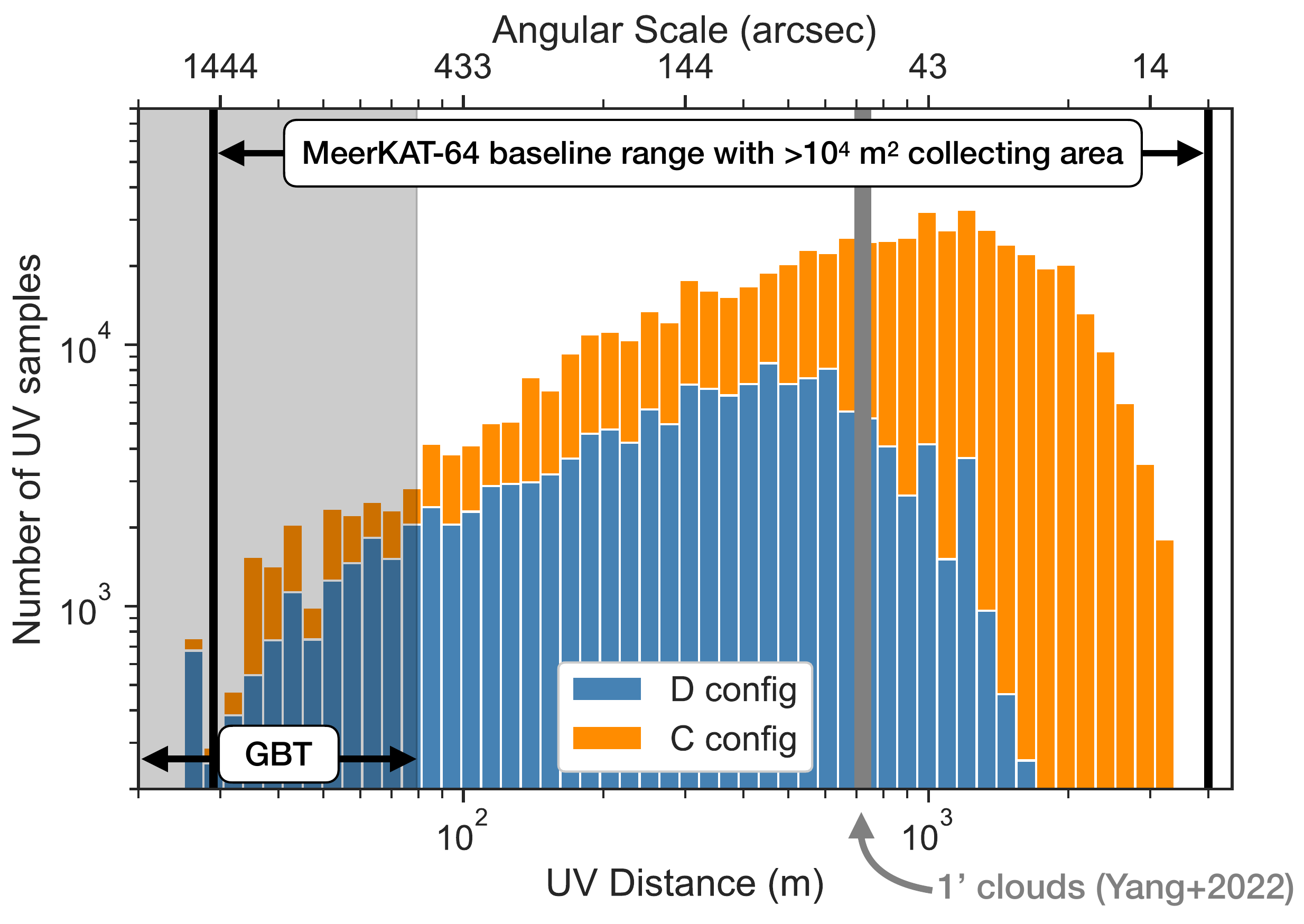}
    \caption{LGLBS $uv$-sampling as a function of $uv$ distance. We show a stacked histogram of the C and D configuration representing the sampling for a single \hi\ channel. The shaded gray area shows the scales that the GBT is sensitive, adopting the GBT beam model from \citet{pingel2018}. The range indicated by the black vertical lines shows where MeerKAT's baseline distribution has an effective collecting area of $>10^4$m$^2$ with; the MeerKAT-16 baseline distribution is similar to the full MeerKAT core, though with $\sim10\times$ less collecting area from fewer baselines. The vertical gray line is the typical $1\arcmin$ scale from the clouds identified in \citet{Yang2022}. This demonstrates that LGLBS, and the full MeerKAT array, have sufficient $uv$-sampling on the scales relevant to recover the \citet{Yang2022} extraplanar emission. The lack of detection in both LGLBS and MeerKAT-64 suggests a systematic issue with the MeerKAT-16 observations.}
    \label{fig:wlm_c+d_hi_uvdist}
\end{figure}

To explore whether the LGLBS data sample the relevant angular scales to detect the \citet{Yang2022} clouds, Figure \ref{fig:wlm_c+d_hi_uvdist} shows the $uv$-samples in the VLA C and D configurations for our WLM observations and compares to the $1\arcmin$ diameter of the \cite{Yang2022} clouds.
The figure shows that $1\arcmin$ scales are well-sampled in our data and thus should be sensitive to the full spatial range of MeerKAT-16 emission.

We further show in Figure \ref{fig:wlm_c+d_hi_uvdist} that the MeerKAT-64 observations from \citet{Kolhe2026} also sample this full range of angular scales.
We use the \texttt{ska\_ost\_array\_config} package\footnote{\url{gitlab.com/ska-telescope/ost/ska-ost-array-config}} and find that the range of baselines where MeerKAT-64 has $>10^4$~m$^2$ effective collecting area is well-matched to our VLA C and D baseline range, and indeed exceeds the VLA's short baseline coverage due to the smaller dish diameters and dense core of MeerKAT.
We note that, using the same approach above for the MeerKAT-16 dishes identified in the archive metadata for the \citet{Ianjamasimanana2020} observations, the MeerKAT-16 sample a similar range just with $\sim10\times$ less collecting area.
Since the MeerKAT-64 observations had more time on source than the \citet{Ianjamasimanana2020} MeerKAT-16 observations, the \citet{Kolhe2026} observations should have significantly more sensitivity on $1\arcmin$ scales to recover the clouds reported in \citet{Yang2022}.

\section{Discussion} \label{sec:discussion}
The non-detection of ram-pressure-stripped \hi\ toward WLM has important implications for the IGM. 
For example, \citet{Yang2022} showed that the presence of ram-pressure-stripped clouds toward WLM implied a surprisingly high IGM density, $n_{\mathrm{IGM}}\geq 5.2\times10^{-5}~\percc$, given the surface density of the ISM and the total mass in WLM \citep[see][]{Gunn1972}. 
Because we do not detect such ram-pressure-stripped \hi, we do not need to invoke such IGM densities, which were significantly higher than estimates of $n_{\mathrm{IGM}}$ measured in the M81 group \citep[$n_{\mathrm{IGM}}\geq 4\times10^{-6}~\percc$;][]{McConnachie2007} and the Local Group \citep[$n_{\mathrm{IGM}}\sim10^{-6}$--$10^{-5}~\percc$;][]{BureauCarignan2002}.
We note, though, that these results do not rule out significant ram pressure stripping altogether in this direction. We place important limits on the amount of ram-pressure-stripped neutral gas trailing WLM, but we cannot rule out the presence of stripped ionized gas in this direction.

\begin{figure}
    \centering
    \includegraphics[width=\linewidth]{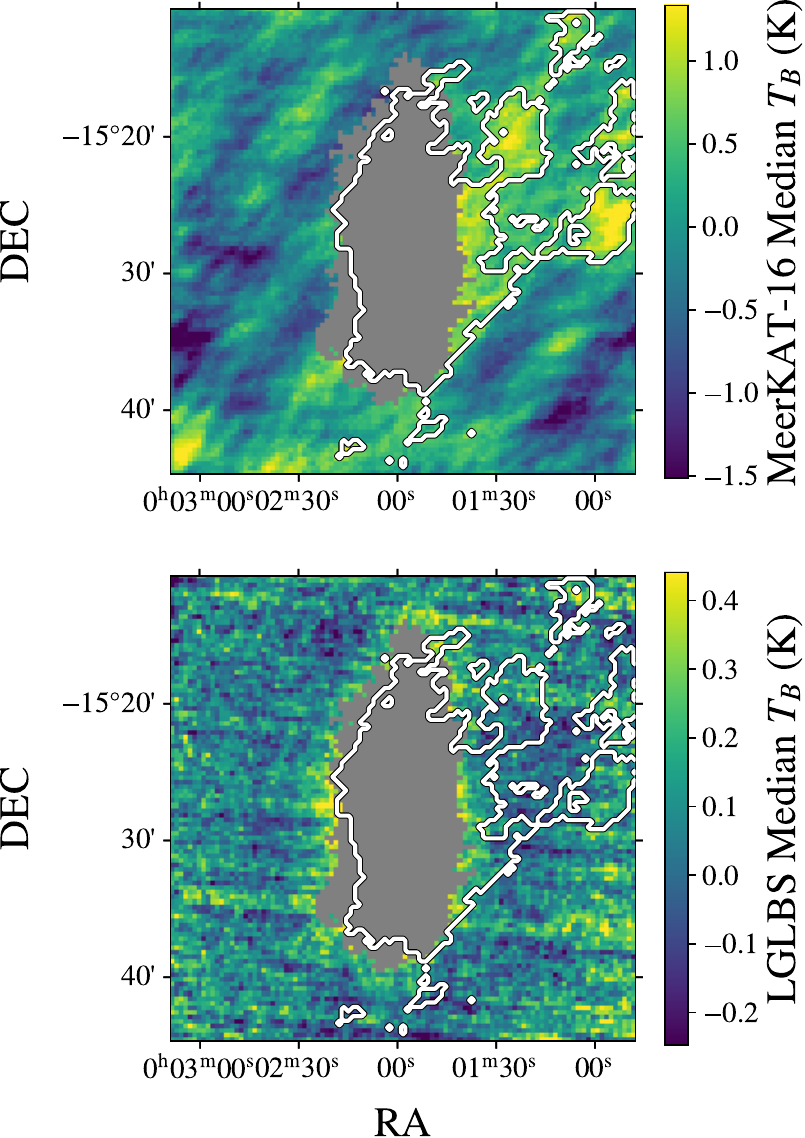}
    \caption{The median brightness temperature of the spectrum in each pixel of the 60\arcsec map from \citet{Ianjamasimanana2020} (top) and LGLBS (bottom). Contours of the extended emission reported by \citet{Yang2022} are overlaid in white. Striping --- likely associated with RFI --- is evident in the \citet{Ianjamasimanana2020} map, and the detections reported by \citet{Yang2022} are clearly associated with the positive-valued peaks in these residual features. Pixels where we detect significant \hi\ emission in the LGLBS cube are masked out in both maps.}
    \label{fig:striping}
\end{figure}

From the lack of detection in the LGLBS VLA data and the full MeerKAT array from \citet{Kolhe2026}, 
we suggest that the \citet{Yang2022} reported detections probably result from residual emission in the \citet{Ianjamasimanana2020} maps, likely caused by unflagged RFI. In Figure \ref{fig:striping}, we highlight the presence of the striping that likely results from such RFI and the association of positive-valued stripes with the clouds reported by \citet{Yang2022}.
The differing results from our work, \citet{Yang2022}, and \citet{Kolhe2026} highlight the value in having independent \hi\ data sets rigorously test the detection of unexpected extended emission.
Direct comparisons between archival and new interferometric data, within the range they are both sensitive to, will be a valuable check for automated data calibration and imaging planned for the next generation of radio interferometers.

In an upcoming paper, F. Caballero Vargas et al. (in preparation) will present an additional analysis of LGLBS WLM \hi\ observations that incorporate VLA A- and B-configuration observations covering the central pointing of the larger mosaic we show here. They will also focus on potential \hi\ outflows and the lagging diffuse \hi. Similar to the findings we present here, though, the initial results at finer spatial scales also do not reveal a population of ram-pressure-stripped \hi\ clouds consistent with the detections reported in \citet{Yang2022}.

\begin{acknowledgments}
We thank Roger Ianjamasimanana for kindly providing MeerKAT-16 \hi\ cube of WLM.  
We also thank the anonymous reviewer for their valuable comments and insights.
This research was supported by the National Science Foundation awards 2205628, 2205629, 2205630, and 2205631. 
The National Radio Astronomy Observatory and Green Bank Observatory are facilities of the National Science Foundation operated under cooperative agreement by Associated Universities, Inc.
D.R.R. was supported by a National Science Foundation Astronomy and Astrophysics Postdoctoral Fellowship under award AST-2303902.
F.C.V. acknowledges support from the NSF under Cooperative Agreements No. 1647375 and 1647378, including the Radio Astronomy Data Imaging and Analysis Lab (RADIAL) Research \& Training Experience program.
\end{acknowledgments}

\facilities{VLA, GBT}

\software{astropy \citep{2013A&A...558A..33A,2018AJ....156..123A,2022ApJ...935..167A},  
         \texttt{ska\_ost\_array\_config} package (\url{gitlab.com/ska-telescope/ost/ska-ost-array-config}),
         radio-astro-tools (spectral-cube, radio-beam) \citep{SPECTRALCUBE2020}}

\bibliography{refs}{}
\bibliographystyle{aasjournalv7}

\end{document}